\documentstyle[12pt]{article}
\newcommand{\eq}[1]{Eq.~(\ref{#1})}

\newcommand{\nl}{\nonumber \\}

\def\beq{\begin{equation}}
\def\eeq{\end{equation}}
\def\beqa{\begin{eqnarray}}
\def\eeqa{\end{eqnarray}}

\begin{document}

\rightline{DFTT 17/95}
\rightline{NORDITA 95/13-P}
\rightline{\hfill February, 1995}

\vskip 1cm

\centerline{\Large \bf Gauge theory renormalizations}
\centerline{\Large \bf from the open bosonic string}

\vskip 1cm

\centerline{\bf Paolo Di Vecchia,
                   \footnote{e-mail: DIVECCHIA@nbivax.nbi.dk} }
\centerline{\sl NORDITA}
\centerline{\sl Blegdamsvej 17, DK-2100 Copenhagen \O, Denmark}

\vskip .3cm

\centerline{\bf Alberto Lerda,}
\centerline{\sl Dipartimento di Fisica Teorica, Universit\`a di Salerno}
\centerline{\sl I-84100 Salerno, Italy}
\centerline{\sl and I.N.F.N., Sezione di Napoli}

\vskip .3cm

\centerline{\bf Lorenzo Magnea,}
\centerline{\sl Dipartimento di Fisica Teorica, Universit\`a di Torino}
\centerline{\sl Via P.Giuria 1, I-10125 Turin, Italy}
\centerline{\sl and I.N.F.N., Sezione di Torino}

\vskip .3cm

\centerline{\bf Raffaele Marotta \footnote{Work partially supported by
the Italian Ministero degli Affari Esteri} }
\centerline{\sl Dipartimento di Scienze Fisiche, Universit\`a di Napoli}
\centerline{\sl Mostra D'Oltremare, Pad. 19, I-80125 Napoli, Italy}
\centerline{\sl and I.N.F.N., Sezione di Napoli}

\vskip 1cm

\begin{abstract}

We present a unified point of view on the different methods available in
the literature to extract gauge theory renormalization constants from
the low-energy limit of string theory. The Bern-Kosower method, based on
an off-shell continuation of string theory amplitudes, and the
construction of low-energy string theory effective actions for gauge
particles, can both be understood in terms of
strings interacting with background gauge fields, and thus reproduce,
in the low-energy limit, the field theory results of the background
field method. We present in particular a consistent off-shell
continuation of the one-loop gluon amplitudes in the open bosonic string
that reproduces exactly the results of the background field method in
the Feynman gauge.

\end{abstract}

\newpage

\section{Introduction}

The interest in extracting from string theory information about the
effective interactions of the low-energy degrees of freedom is as
old as string theory itself.

Using a variety of approaches several authors~
\cite{mets}-\cite{berkos} have studied
how to extract one-loop field theoretical quantities, such as scattering
amplitudes, renormalization constants and $\beta$ functions, from ultraviolet
finite string theories.

In particular, it has become clear through the work of Bern and Kosower
that, whether or not string theory is a fundamental theory of nature, it is a
valuable tool to obtain results in quantum field theory that would have
been very hard to get otherwise. They extracted from string theory the
QCD $\beta$ function~\cite{berkosbet}, and, more generally, they
used string theory to generate rules to calculate
one-loop amplitudes in QCD~\cite{berkos}. Using these rules, they were able to
calculate all the helicity amplitudes involving five gluons at one loop
\cite{fiveglu}, a remarkable achievement from the point of view of
traditional techniques. Similar methods were also applied to gravity
\cite{berdun}.

Other approaches are based on the study of strings interacting with
external backgrounds~\cite{kaplu,fratse1}.
With this method one can compute from string theory the running of the
low-energy couplings, determine threshold corrections in
string-inspired grand unified theories~\cite{kaplu}, and obtain several
other low-energy quantities of interest~\cite{kaplu2}.

The main obstacle to the extraction of this kind of information lies
in the fact that present day string technology is essentially restricted
to the calculation of on-shell quantities, whereas renormalization in
field theory is best carried out for general, off-shell Green's functions.

This obstacle can be circumvented, however, either by attempting an
off-shell continuation of the string theory amplitudes calculated
with traditional techniques \cite{berkosrol}, or by evaluating
low-energy effective actions in the presence of background fields
satisfying the equations of motion, as done in Ref.~\cite{kaplu}.

The purpose of this letter is to establish a connection between the two
methods, and to show how they are both strictly related, in the
field theory limit, to the background field method. In the
process, we believe we will clarify several issues concerning the
relationship between string theory amplitudes and renormalizations, and
we will show the consistency, from the point of view of the gauge
theory, of a particular prescription for the off-shell continuation of
the string amplitudes, different from the one given
in Ref.~\cite{berkosrol}.

We must emphasize that we are not going to require
the overall consistency of the string theory we work with. Rather,
we consider string theory as an efficient bookkeeping device that may
generate economically field theoretical results, in the spirit of Bern
and Kosower's approach.
Thus we work with the open bosonic string, and we formally continue
the consistent on-shell amplitudes off shell, and to arbitrary
space-time dimension. We will see at the end that, having taken
correctly the field theory limit, and
decoupling the $1/\alpha'$ tachyonic
divergences, we will recover consistent field theory results.

We will briefly discuss how gauge theory
renormalizations are related to string theoretical amplitudes and
effective actions. We will then consider in some detail the two-gluon
amplitude in the open bosonic string, restricting our attention to the
$SU(N)$ gauge group, and we will show how, in the field theory
limit, it coincides precisely with the amplitude calculated in the
background field Feynman gauge, off shell and with dimensional
regularization. We will argue that the prescription used to extend
off shell the two-gluon amplitude can be applied to other amplitudes as
well, and it gives consistent renormalizations for the gauge theory,
obeying the appropriate Ward identities. We will then examine the
one-loop effective action for the open bosonic string, in a
slowly varying Yang-Mills background, and we will show that it can be
interpreted in practice as an infrared limit of the previously
calculated two-gluon amplitude. Also in this case we can establish a
precise connection with the corresponding calculation in field theory,
which is again carried out with the background field method, but this
time regularized with a proper time cutoff.

To lay the grounds for an
extension of the present calculations to higher loops, we will use from
the beginning the $h$-loop operator formalism discussed in
Ref.~\cite{copgroup}, but
the reader is advised that only one-loop calculations will be performed,
and it is straightforward to map the present notation into the more
standard one of, say, Ref.~\cite{GSW}. Several technical points and the
details of the calculations are left to a subsequent publication
\cite{usbig}.

\section{A note on renormalization constants}

String theory tells us how to compute on-shell scattering amplitudes
among physical string states. In field theory, on the other hand,
we know through the
reduction formulas that on-shell scattering amplitudes are given by
the on-shell limit of the truncated Green's functions of the theory,
multiplied by the appropriate powers of the residues of the propagators.

When we calculate one-loop string theory amplitudes, and take the
field theory limit, generating the ultraviolet divergences of
the low-energy effective theory, we are thus calculating
{\sl unrenormalized}, on-shell, truncated Green's functions.

This is a
problem if the low-energy theory contains gauge bosons, with the
associated mass (infrared and collinear) divergences, because
these Green's functions are
ill-defined even in field theory, due to the possible cancellation of
mass and ultraviolet divergences. This is apparent
if one uses dimensional regularization: in this case the Green's
function receives contributions from diagrams in which the loop is
isolated on an external leg, and since such a loop integral is defined
to vanish in dimensional regularization, the Green's function suffers,
{\sl even in field theory}, from a $0/0$ ambiguity.

It is easy to deal with this ambiguity in field theory: one simply
calculates the Green's function off shell, thus getting rid of
mass divergences, and subsequently renormalizes it, subtracting out
the ultraviolet divergences that are left. Finally, one takes the
on-shell limit, and recovers a purely infrared divergence, which is a
genuine contribution to the $S$-matrix, and will be cancelled when
forming a sufficiently inclusive cross section.

In string theory the situation is complicated by the fact that there is
no obvious consistent way of continuing the scattering amplitudes
off shell, thus disentangling mass from ultraviolet divergences
in the field theory limit. The amplitudes at one loop
are inherently ambiguous, so one has to choose a
prescription to regularize this ambiguity, and only a posteriori one can
verify the consistency of the calculation \cite{berkosrol}.

Assuming that one such prescription has been chosen, the computation of
renormalization constants is relatively straightforward; one just has to
keep in mind the relationship between truncated and one-particle
irreducible Green's functions (proper vertices). To set our notations,
consider a pure gauge theory, denote by
$\Gamma_M(p_1, \ldots , p_{M-1})$ the
proper $M$-point vertex, and by $W_M(p_1, \ldots , p_{M-1})$
the truncated connected $M$-point
Green's function, with momentum conservation already taken into account.
Let $Z_A$, $Z_3$ and $Z_4$ be the multiplicative
renormalization constants associated with the gluon wave function and
with the three- and four-point vertices respectively. Then the
relation between the first few truncated functions and the corresponding
proper vertices is
\beqa
W_2(p) & = & - i \left(\Gamma_2(p)\right)^{-1}
             \left(\Gamma_2^{(0)}(p)\right)^2~~, \nl
W_3(p_1,p_2) & = & i \Gamma_3(p_1,p_2) \prod_{k=1}^3
   \left[\left(\Gamma_2(p_k)\right)^{-1} \Gamma_2^{(0)}(p_k)\right]~~, \\
W_4(p_1,p_2,p_3) & = & i \left[ \Gamma_4(p_1,p_2,p_3) -
   \sum_{(i,j)} \Gamma_3(p_i,p_j) \left(\Gamma_2(p_i + p_j)\right)^{-1}
      \Gamma_3(p_k,p_l) \right] \nl
  & & \prod_{k=1}^4
   \left[\left(\Gamma_2(p_k)\right)^{-1} \Gamma_2^{(0)}(p_k)\right]~~,
     \nonumber
\label{wgamma}
\eeqa
where $i \left(\Gamma_2^{(0)}(p)\right)^{-1}$ is just the free
propagator, and the sum in the last equation is over the three channels
of the four point amplitude. As a consequence, the truncated Green's
functions must be renormalized according to
\beqa
W_2 & = & W_2^{(R)} Z_A~~,\nl
W_3 & = & W_3^{(R)} Z_3^{-1} Z_A^3~~, \\
W_4 & = & W_4^{(R)} Z_4^{-1} Z_A^4~~,  \nonumber
\label{itself}
\eeqa
where the Ward identity $Z_4 = Z_3^2 Z_A^{-1}$ has been used.

The ultraviolet divergences encountered in the field theory limit
of a string amplitude are thus related to the combinations of
renormalization constants given by Eq.~(2).
Having established this, we go on to the calculation of the two-gluon
amplitude, using the open bosonic string.

\section{The two-gluon amplitude}

To compute the $M$-gluon, $h$-loop amplitude $A_M^{(h)}$ in the open bosonic
string in $d$ dimensions, one can use the operator formalism discussed in
Ref.~\cite{copgroup}, {\it i.e.} one can saturate the $M$-point $h$-loop
vertex $V_{M;h}$ with gluon states of momenta $p_i^\mu$,
polarizations $\varepsilon_i^\mu$, and color indices $a_i$.
The field theory limit of $A_M^{(h)}$ yields the $h$-loop contribution
to the on-shell, connected, truncated Green's functions $W_M$ discussed
in the previous section.

For $M = 2$, it is apparent that the result is ill-defined on shell,
since all the kinematical invariants containing the gluon momenta vanish
on shell. Furthermore, an ambiguity persists for any $M$ \cite{berkosrol},
and it is twofold: on the one hand, if the external states are kept on
the mass shell, the field theory limit contains ratios of vanishing
invariants, and is thus ambiguous, not unlike the unrenormalized on-shell
truncated Green's functions $W_M$.
On the other hand, if one attempts a na\"{\i}ve continuation off shell,
the world-sheet projective invariance of the amplitude is broken, and
the result acquires a spurious dependence on the local coordinates of
the punctures. In fact, the $M$-point vertex $V_{M;h}$ depends on the
Koba-Nielsen variables $z_i$ through $M$ projective transformations
$V_i(z)$, satisfying the constraint $V_i(z_i) = 0$. The on-shell
amplitude is invariant under changes in the $V_i$'s, but the off-shell
continuation is not. This is again not unlike the situation encountered
in the gauge theory, where the on-shell limit is gauge invariant,
but the off-shell continuation is not.

This twofold ambiguity was solved in \cite{berkosrol} with a particular
choice of the $V_i$'s, leading to non-ambiguous results in almost all
cases. That choice turns out to be equivalent to the request that the
gluon wave-function renormalization vanish.

Here we propose a different solution of the ambiguity, and we show that,
with our prescription, the field theory limit is consistent and
unambiguous, and it corresponds to the background field Feynman gauge.
Our prescription is simply to treat the freedom to choose the $V_i$'s as
a gauge freedom, which we fix to achieve the maximum simplification in
the calculations. Our choice, at one loop, is
\beq
V_i(z) = z_i (z - z_i)~~,
\label{vi}
\eeq
while at $h$ loops it may be necessary to choose $V_i$'s depending on
the moduli of the surface.
Next we continue the momenta off the mass shell with the simplest choice
preserving Bose symmetry, namely we set
\beq
p_i^2 = m^2~~,~~~~\forall i~~,
\label{pi}
\eeq
while we keep the transversality condition $\varepsilon_i \cdot p_i = 0$
to simplify the calculation.
Relaxing this condition would generate more
cumbersome expressions but would not change our results.
In string theory, the choice of \eq{pi} is suggested by the observation
\cite{kounnkir} that it is possible to construct consistent string
theories in which the entire spectrum is shifted by a fixed amount
$m^2$, by considering curved space-times, and in particular certain
coset spaces. Since we are interested in the field theory limit, and
specifically in ultraviolet-dominated quantities, such as
renormalization constants, we can simply take the curvature of
space-time as an infrared regulator, and we need not worry about global
properties.

Having established our prescription, let us start
by writing down the two-gluon, $h$-loop correlation function
derived using the operator formalism. It is
\beqa
A_2^{(h)}&=&C_h~{\cal N}_0^2
{}~\int [dm]_h ~\exp\big[p_1\cdot p_2
{}~{\cal G}(z_1,z_2)\big] \left[V_1'(0) V_2'(0)
\right]^{- (p_1 \cdot p_2)/2} \nl
& & \Big[\varepsilon_1\cdot\varepsilon_2
{}~\partial_{z_1}\partial_{z_2}{\cal G}(z_1,z_2)
+\varepsilon_1\cdot p_2~\varepsilon_2\cdot p_1
{}~\partial_{z_1}{\cal G}(z_1,z_2)~\partial_{z_2}{\cal G}(z_1,z_2)
\Big]~~,
\label{2gluhloop}
\eeqa
where $z_1$ and $z_2$ are the real Koba-Nielsen
variables associated with the two gluons, $[dm]_h$ is the $h$-loop
measure on moduli space
for the open bosonic string \cite{copgroup}, and ${\cal G}(z_1,z_2)$
is the world-sheet bosonic Green's function at $h$ loops.
$C_h$ is the $h$-loop vertex normalization factor containing the
$d$-dimensional Yang-Mills coupling constant
$g_d$, the inverse string tension
$\alpha'$, and the appropriate $SU(N)$ Chan-Paton factor, while
${\cal N}_0$ is the normalization of the gluon state. We have omitted a
momentum conservation delta function, and we are measuring momenta in
units of $(2 \alpha')^{-1/2}$. The amplitude
$A_2^{(h)}$ corresponds to a planar string diagram with $h$ loops, and
two punctures on one boundary.

\eq{2gluhloop} is a remarkable expression, as any field theorist
will realize. It generates, in the field theory limit, all the loop
corrections to the Yang-Mills two-point function, something for which no
closed expression exists in field theory. In particular, since we will
show that in the field theory limit one recovers the background field
method, \eq{2gluhloop} is a master formula
containing all the information
necessary to determine the multi-loop Yang-Mills $\beta$ function.

Specializing to one loop ($h=1$), the measure on moduli space
will depend on five variables: the two Koba-Nielsen variables $z_1$ and
$z_2$, and the three
moduli of the projective transformation that defines the annulus, which
can be chosen (in the Schottky parametrization) as the multiplier $k$
and the two fixed points $\eta$ and $\xi$ \cite{copgroup}.
Three of these variables
can be fixed using projective invariance. Our choice is to
fix $\eta = 0$, $z_1 = 1$, and $\xi = \infty$.
Then the integration region for the remaining moduli
$z_2$ and $k$ can be determined by going back to the sewing procedure
used to construct the one-loop two-point vertex from the tree-level
four-point one \cite{cop1}. It is given by
\beq
1\geq z_2 \geq k \geq 0~~.
\label{region}
\eeq
With these choices, the modular measure at one loop becomes,
\beq
[dm]_1 = dz_2~dk~ {1\over k^2} \left(-{1\over {2\pi}}\log k\right)^{-d/2}
\prod_{n=1}^\infty (1-k^n)^{2-d}~~.
\label{modmes}
\eeq
Notice that the multiplier $k$ is related to the usual modular parameter
${\tilde \tau}$ by
\beq
k = {\rm e}^{2\pi{\rm i}\,{\tilde \tau}}~~.
\label{ktau}
\eeq
Below, we will use instead of ${\tilde \tau}$, which is purely
imaginary, the real variable $\tau = - {\rm i} \pi {\tilde \tau}$, which
is integrated between $0$ and $\infty$. Similarly, instead of the
Koba-Nielsen variables $z_i$, we will use the real variables
$\nu_i = - \frac{1}{2} \log z_i$, so that with our choices
$\nu_1 = 0$ while $\nu_2$ is integrated between $0$ and
$\tau$~\cite{berbos}.

The bosonic Green's function ${\cal G}(z_1,z_2)$ at one loop is given by
\beq
{\cal G}(z_1,z_2) =
\log \left[ - 2\pi{\rm i}~\frac{\theta_1\left(\frac{{\rm i}}{\pi}
(\nu_2 - \nu_1) \Big| \frac{{\rm i}}{\pi} \tau \right)}{\theta'_1
\left(0 \Big| \frac{{\rm i}}{\pi} \tau \right)} \right] -
\frac{(\nu_2-\nu_1)^2}{\tau} - \nu_1 - \nu_2~~,
\label{green}
\eeq
where $\theta_1$ is the Jacobi $\theta$-function.
\eq{green} is related to the more commonly used one-loop correlation
function $G(\nu)$ of, say, Ref.~\cite{GSW}, by
\beq
G(\nu_2 - \nu_1) = {\cal G}\left(z_1(\nu_1),z_2(\nu_2)\right)
      + \nu_1 + \nu_2~~.
\label{gswgreen}
\eeq
Our choice of the $V_i$'s, \eq{vi}, is made precisely
to cancel the factors arising from $\nu_1+\nu_2$
in the exponent of \eq{2gluhloop}.

We are now ready to write down explicitly the one-loop version of
\eq{2gluhloop}. Following the strategy of Ref.~\cite{berkos}, we first
integrate by parts the term containing a double derivative of the
Green's function. Then we mimick dimensional regularization by formally
taking the dimension of space-time to be $d = 4 - 2 \epsilon$. Finally
we write down explicitly all the numerical factors, and reinstate
$\alpha'$ in preparation for the field theory limit, $\alpha'
\rightarrow 0$. The result takes the form
\beq
A_2^{(1)} = {N\over 2}~\frac{g^2}{(4 \pi)^2}~\delta_{ab}~
(4 \pi \mu^2 \cdot 2 \alpha')^{2 -d/2}~
(\varepsilon_1\cdot\varepsilon_2~p_1\cdot p_2 - \varepsilon_1\cdot p_2~
\varepsilon_2\cdot p_1 )~R(p_1 \cdot p_2)~~.
\label{2glu1loop}
\eeq
Here we wrote $g_d = g \mu^\epsilon$ in accordance with field theory
conventions, and we
have chosen not to enforce transversality and momentum conservation just
yet, to display the correspondence with the field theory off-shell
amplitude. The relevant factor is the integral
$R(p_1 \cdot p_2)$, defined by
\beq
R(s) \equiv \int_0^\infty \!\!d\tau \int_0^\tau \!\!d\nu
\left[{\rm e}^{2 \tau}~(\tau)^{- d/2}
{}~\prod_{n=1}^\infty \left(1-{\rm e}^{-2 n \tau}\right)^{2 - d}
{}~{\rm e}^{~2 \alpha' s \, G(\nu)}~\left(\partial_\nu G(\nu)\right)^2
\right]~~.
\label{rint}
\eeq
\eq{rint} is the starting point to take the field theory
limit, which in these coordinates corresponds to the limit $k = {\rm
e}^{- 2 \tau} \rightarrow 0$, or $\tau \rightarrow \infty$, as in
\cite{berkos,kaj}. In this limit the Green's function behaves as
\beq
G(\nu) = - \frac{\nu^2}{\tau} + \log\left(2 \sinh(\nu)\right) -
         4 {\rm e}^{- 2 \tau} \sinh^2(\nu) + O({\rm e}^{- 4 \tau})~~,
\label{limG}
\eeq
and it is convenient to change variables to ${\hat \nu} = \nu/\tau$.
Substituting \eq{limG} into \eq{rint}, and keeping only the terms that
are finite in the limit $k = 0$ (the divergent terms correspond to the
infrared divergence associated with the tachyon), we find
\beq
R(s) = \int_0^\infty \!\!d\tau \int_0^1 \!d{\hat \nu}~
\tau^{1 - d/2}~{\rm e}^{2 \alpha' s ({\hat \nu} - {\hat \nu}^2)
\tau} \left[(1 - 2 {\hat \nu})^2 (d - 2) - 8\right]~~.
\label{rintlim}
\eeq
The integrals are now elementary, and the answer is, reverting to the
dimensional continuation parameter $\epsilon$,
\beq
R(s) = - 2~\Gamma(\epsilon)~(-2 \alpha' s)^{- \epsilon}~
\frac{11 - 7 \epsilon}{3 - 2 \epsilon}~B(1 - \epsilon,1 - \epsilon)~~.
\label{rfin}
\eeq
Putting everything together we find that the two-gluon one-loop
amplitude is
\beqa
A_2^{(1)} & = & - N \delta_{ab}~\frac{g^2}{(4 \pi)^2}~
\left(\frac{4 \pi \mu^2}{-p_1 \cdot p_2}\right)^\epsilon~
(\varepsilon_1\cdot\varepsilon_2~p_1\cdot p_2 - \varepsilon_1\cdot p_2~
\varepsilon_2\cdot p_1) \times \nl
& \times & \Gamma(\epsilon)~\frac{11 - 7 \epsilon}{3 - 2 \epsilon}~
B(1 - \epsilon,1 - \epsilon)~~.
\label{fin2glu}
\eeqa
This result is exactly equal, including the normalization and to all
orders in $\epsilon$, to the result one obtains calculating the
gluon vacuum polarization in field theory, using the background field
method and the background field Feynman gauge, with Feynman rules given
for example in Ref. \cite{abbott} (the sign of $p_1 \cdot p_2$
is due to the fact that we are working with the metric of Ref. \cite{GSW}).
One is lead to the conclusion (which we
will verify in the last section) that this is the gauge to which our
prescription leads, as might have been expected from the on-shell
analysis of Ref.~\cite{berfield}.
In particular, the divergence in \eq{fin2glu} must
be cancelled by a wave function renormalization, and using the
background field method Ward identity $Z_g = Z_A^{-1/2}$ and minimal
subtraction one recovers the correct Yang-Mills $\beta$ function.

\section{Bosonic string in an external background}

In this section we will follow a different approach, which also leads to
a correct determination of the Yang-Mills $\beta$ function. We will study an
open bosonic string in interaction with a slowly varying external
non-abelian field, and we will extract the one-loop contribution to the two
point function. In the one-loop effective action this corresponds
to a renormalization of the gauge field classical action, {\sl i.e.}
to a wave function renormalization.

Let us consider the partition function of an open bosonic string
interacting with an external non-abelian $SU(N)$ background, pointing
in a specific direction in colour space, and let us denote it
by $Z(F_{\mu \nu})$. We write the string coordinate $X^\mu(z)$ as
\beq
X^\mu(z) = \frac{x^\mu}{\sqrt{2 \alpha'}} + \xi^\mu(z)~~,
\label{xmu}
\eeq
so that $X^\mu$ and $\xi^\mu$ are dimensionless, while the zero mode $x^\mu$
has dimensions of length; then the planar contribution
to $Z(F_{\mu \nu})$ is given by
\beqa
Z_{pl}(F_{\mu \nu}) & = & \sum_{h=0}^{\infty} N^h g_s^{2 h -2}
\int\frac{d^d x}{(2 \alpha')^{d/2}} \int_h D\xi Dg~
{\rm e}^{- S_0(\xi,g;h)} \times \nl
& \times & Tr \left[ P_z \exp \left( - i \alpha' g_d F_{\mu \nu}(x)
\int_h dz ~\partial_z \xi^{\mu} (z)  \xi^{\nu} (z)
\right) \right]~~.
\label{part}
\eeqa
Here the symbol $P_z$ reminds us that in the open string the $z$ variables
along the world-sheet boundary are ordered; $g_s$ is the dimensionless
string coupling constant, related to $g_d$ through
\beq
g_s = \frac{g_d}{\sqrt{2}}\,(2 \alpha')^{1 - d/4}~~;
\label{gstring}
\eeq
$F_{\mu \nu}$ is a matrix proportional to one of the $SU(N)$
generators $\lambda_a$ (normalized by ${\rm Tr}(\lambda_a \lambda_b) =
\frac{1}{2} \delta_{ab}$), and we take it to be
independent of $\xi^\mu$, and slowly varying
as a function of the zero mode $x^\mu$; finally, the classical action on
a genus $h$ manifold with metric $g_{\alpha \beta}$ is
\beq
S_0(\xi,g;h) = \frac{1}{2 \pi}
\int_h d^2 z ~\sqrt{g} g^{\alpha \beta} \partial_{\alpha} \xi
\cdot  \partial_{\beta} \xi~~.
\label{classact}
\eeq

We are interested in computing the contribution to \eq{part} which is
quadratic in the field strength, and which shall be denoted by
$Z_2(F_{\mu \nu})$. It is given by
\beqa
Z_2(F_{\mu \nu}) & = & \sum_{h=0}^{\infty} N^h g_s^{2 h}
\left[ - \frac{1}{4} \int d^d x ~F_{\mu \nu}^a(x) F_{\rho \sigma}^a(x) \right]
\times \nl
& \times & \frac{1}{2} \int dz dw \left\langle
\partial_z \xi^\mu(z) \xi^\nu(z) \partial_w \xi^\rho(w) \xi^\sigma(w)
\right\rangle_h~~,
\label{z2corr}
\eeqa
where $a$ is a fixed index, and the correlator is defined by the functional
integral in \eq{part}, formally continued to arbitrary $d$.
The contribution from the disk ($h = 0$) is just the classical gauge action
for the background field,
\beq
Z_{2}^{(0)} = - \frac{1}{4} \int d^d x ~
F_{\mu \nu}^{a} F_{\mu \nu}^{a}~~.
\label{zdisk}
\eeq
The contribution from the annulus ($h=1$) has the same form, but
it is multiplied by a non-trivial integral,
\beq
Z_{2}^{(1)} = - \frac{1}{4} \int d^d x ~F_{\mu \nu}^{a} F_{\mu \nu}^{a}~
{\hat{Z}}_2~~,
\label{zannulus}
\eeq
where
\beq
{\hat{Z}}_2  =  \frac{N}{2} \frac{g_d^2}{(2 \pi)^d}\,
(2 \alpha')^{2 - d/2}
\int \frac{[dm]_1}{z_2} \lim_{z_1 \rightarrow 1}
\left[ \partial_{z_1} {\cal G} (z_1, z_2) - \frac{1}{2 z_1} \right]^2~~,
\label{zfactor}
\eeq
while ${\cal G}(z_1, z_2)$ is the bosonic Green's function given in
\eq{green}, and $[dm]_1$ is the measure given in \eq{modmes}.

Introducing the variables $\nu$ and $\tau$ defined below \eq{ktau},
we can rewrite the expression in \eq{zfactor} as
\beqa
{\hat{Z}}_2 & = & \frac{N}{2} \frac{g_d^2}{(4 \pi)^{d/2}}\,
(2 \alpha')^{2 - d/2}
\int_{0}^{\infty} d \tau ~ {\rm e}^{2 \tau }
\left(\tau\right)^{- d/2}
\prod_{n=1}^\infty \left(1-{\rm e}^{-2 n \tau}\right)^{2 - d}
\int_{0}^{\tau} d \nu
\left[\partial_\nu G (\nu) \right]^2   \nl
& = & \frac{N}{2} \frac{g^2}{(4 \pi)^2}
\left(4 \pi \mu^2 \cdot 2 \alpha'\right)^{2 - d/2} R(0)~~,
\label{newzfact}
\eeqa
where $R(s)$ was defined in \eq{rint}.

We see that we recover precisely the expression of \eq{2glu1loop}, but
with the infrared cutoff $p_1 \cdot p_2 = - m^2$ removed inside the
integral. The reason is that the calculation leading to \eq{2glu1loop}
can also be understood as a background field calculation, but with the
background represented by a plane wave vector potential. Here instead we
have used a vector potential linear in $x$, leading to a constant field
strength. The present calculation is thus the long-wavelength
approximation of the previous one, where we have retained only terms
linear in the momentum $p$, so that the exponential factor in
$R(p_1 \cdot p_2)$ has been discarded.

Clearly, the elimination of the exponential term, which acted as an
infrared cutoff, will generate contributions from
the massless states, that will survive in the field theory
limit, and will be infrared divergent in the limit $d \rightarrow 4$, as
expected from field theory.

We can see this explicitly if we repeat the procedure leading to
\eq{rintlim}, discarding as before the tachyonic contributions.

Performing the integration over the puncture we get
\beq
{\hat{Z}}_2 = \frac{N}{2} \frac{g_d^2}{(4 \pi)^{d/2}}
\frac{d - 26}{3} (2 \alpha')^{2 - d/2}
\int_{0}^{\infty} d \tau ~\tau^{1 - d/2}~~.
\label{field1}
\eeq

In $d>4$ the integral over $\tau$ is ultraviolet divergent. It can be
regularized by introducing a cutoff $1/(2 \alpha' \Lambda^2)$ at the
lower limit of integration, corresponding to a proper time cutoff
in field theory. Then we get
\beq
{\hat{Z}}_2 = N \frac{g_{d}^{2} }{(4 \pi)^{d/2}} \frac{d - 26}{3}
 \frac{\Lambda^{d-4}}{d-4}~~,
\label{field2}
\eeq
which agrees with the result obtained in Refs.~\cite{mets,fratse2}, and
gives a logarithmic ultraviolet divergence  as $d \rightarrow 4$, as
expected. We have also verified that, if one calculates the two-gluon amplitude
with the background field method and in the background Feynman gauge, as
suggested in the previous section, but regularizing the loop integral by means
of a proper-time cutoff, rather than dimensional regularization, the term
proportional to $\Lambda^{d-4}$ is precisely given by \eq{field2}. This
is not a surprise, but it strengthens the connection between string theory
and the background field method.

When $d = 4$, the ultraviolet divergence in \eq{field1} can still be
regularized as in \eq{field2}, or with dimensional regularization, but
the integral develops an infrared divergence, as is expected from field
theory and from the analogous calculations performed for the closed
string~\cite{kaplu}. One can then reintroduce an infrared cutoff
in exponential form, in analogy with the calculation leading to
\eq{rfin}. Multiplying the integrand in \eq{field1} with a factor
$e^{- 2 \alpha' m^2 \tau}$ we get
\beq
{\hat{Z}}_2 = \frac{N}{2} \frac{g_d^2}{(4 \pi)^{d/2}} \frac{d - 26}{3}~
m^{d - 4}~\Gamma (2 - d/2)~~.
\label{field3}
\eeq
Once again, expanding around four dimensions $(d = 4 - 2 \epsilon)$ and
using $g_d = g \mu^{\epsilon}$, we get the pole contribution
\beq
{\hat{Z}}_2 = - N \frac{g^2}{(4 \pi)^2} \frac{11}{3}~
\frac{1}{\epsilon}~~,
\label{newbet}
\eeq
thus reproducing the results of the previous section and the correct
Yang-Mills $\beta$ function.

Had we chosen from the beginning to compactify some of the string
coordinates, \eq{newbet} would have received contributions also from the
adjoint scalars associated with the compactified dimensions.
In this case \eq{newbet} generalizes to
\beq
{\hat{Z}}_2 = - N \frac{g^2}{(4 \pi)^2} \left[\frac{11}{3} -
\frac{N_s}{6} \right]~\frac{1}{\epsilon}~~,
\label{betscal}
\eeq
where $N_s$ is the number of compactified dimensions, and the number of
scalars, so that we recover the well known result stating that the
$\beta$ function of four-dimensional Yang-Mills theory coupled
with 22 adjoint scalars vanishes.

\section{Consistency checks and conclusions}

In section 2 we have shown that when the two-gluon amplitude at one loop
is suitably extended off shell, it leads in the field theory limit,
\eq{fin2glu}, to the wave function renormalization
\beq
Z_A = 1 + N\,{g^2\over {(4\pi)^2}}~{11\over 3} ~{1\over \epsilon}~~,
\label{za}
\eeq
which agrees with the result of the background field calculation
performed in section 3 and given by \eq{newbet}.
This result suggests that our prescription to continue
the gluon amplitudes off-shell
gives the renormalization constants of the background field method.
To verify this statement we have to check the background field Ward
identity
\beq
Z_g = Z_A^{-1/2}~~,
\label{ward1}
\eeq
where as usual $Z_g=Z_3\,Z_A^{-3/2}$. This requires that we compute the
three-gluon amplitude $A_3^{(1)}$, continue it off-shell according to the same
prescritpion used for the two-point amplitude, and then extract from it the
ultraviolet divergent contribution in the field theory limit.
The latter, according to Eq.~(2), must be given by
\beq
A_{3,{\rm div}}^{(1)} = (Z_3^{-1}\,Z_A^3 - 1)~A_3^{(0)}~~,
\label{z3}
\eeq
so that we can extract from it the vertex renormalization constant $Z_3$, and
thus $Z_g$. Further, we must verify that our prescription corresponds to a
consistent one-loop renormalization prescription for the gauge theory,
respecting gauge invariance. To do this we have to compute the ultraviolet
divergence of the four-point amplitude, and verify that the Ward identity
$Z_4=Z_3^2\,Z_A^{-1}$ is also satisfied, so that the renormalized coupling
appearing in the four-point vertex is
in fact the square of the renormalized three-gluon coupling.

The details of this calculation are left to a subsequent publication
{}~\cite{usbig}; here we simply give a brief summary of our results.

The three-gluon amplitude $A_3^{(1)}$ is an integral over the modulus
of the annulus and three cyclically ordered
punctures $\nu_1$, $\nu_2$ and $\nu_3$, one of which,
say $\nu_1$, is actually fixed to zero. According to~\cite{berkos},
the terms that are divergent in the field theory limit are those which
arise from ``pinching'' together two of the three punctures.
Because of the cyclic order, only two pinchings are possible, namely
\beq
\nu_2 \to \nu_1=0 ~~~~~~~{\rm and}~~~~~~~\nu_3 \to \nu_2~~.
\label{pinch}
\eeq
Each one of these pinchings isolates the loop on an external leg and
gives rise to an expression which has poles associated to all
possible string states that are exchanged in the
intermediate channel. The only term in this expansion that survives in
the field theory limit is the one corresponding to the exchange of an
intermediate gluon. The coefficient of this term contains ratios of vanishing
momentum invariants, which can be given an unambiguous meaning using the
off-shell prescription of \eq{pi}. We explicitly verified~\cite{usbig}
that, when the pinchings are performed and the field theory limit is
taken, the divergent part of three-gluon amplitude at one loop can
be written as
\beq
A_{3,{\rm div}}^{(1)} = (Z_A^2 - 1)~A_3^{(0)}~~,
\label{3p}
\eeq
which implies the background field Ward identity
$Z_3=Z_A$, and thus \eq{ward1}.
Similarly, we verified from the four-gluon amplitude at one loop that
$Z_4=Z_A$, so that also the Ward identity $Z_4=Z_3^2\,Z_A^{-1}$ is
satisfied, and the theory can be renormalized at one-loop respecting gauge
invariance.

The lessons we have learnt can be summarized as follows. Existing calculations
of gauge-theory renormalization constants from string theory are all strictly
related to the background field method. The calculation of the string effective
action in a slowly varying background gauge field can be seen as the long
wavelength limit of the same calculation performed with a plane-wave
background, and this in turn coincides with the amplitude for the scattering
of the appropriate number of gluons. Gluon scattering amplitudes in the field
theory limit can be consistently continued off shell, and to arbitrary
space-time dimension, recovering the background field Feynman gauge in
dimensional regularization. The correspondence persists if one uses a
proper-time cutoff instead of dimensional continuation. We hope that the
strong simplifications of the calculations due to the fact that we are dealing
with such a simple string theory will make it possible to extend the present
results to higher loops.

\vskip 2.5cm
{\large {\bf Acknowledgements}}
\vskip 0.5cm

One of us (P. Di V.) thanks M. Bianchi, F. Fucito and G. C. Rossi for
many useful discussions during several visits in Rome, University of
Tor Vergata. R. M. thanks the Niels Bohr Institute and Nordita for their
kind hospitality. We also thank K. Roland for his partecipation in the
very early stage of this work and for very useful discussions.

\end{document}